\documentclass[%
 aip,
 jmp,%
 amsmath,amssymb,
 reprint,%
]{revtex4-1}

\usepackage{graphicx}
\usepackage{dcolumn}
\usepackage{bm}

\usepackage{dcolumn}
\usepackage{bm}
\usepackage{mathtools}

\usepackage[bottom]{footmisc}

\usepackage{soul}
\usepackage[normalem]{ulem}

\usepackage[font=small,labelfont=bf,justification=raggedright]{caption}

\usepackage{graphicx}

\begin{document}

\title[]{Spontaneous activity emerging from an inferred network model captures complex temporal dynamics of spiking data}

\author{Cristiano Capone*}
\email{cristiano0capone@gmail.com\\ *authors contributed equally}
\affiliation{Istituto Superiore di Sanit\`{a}, Rome, Italy}
\affiliation{INFN, Sezione di Roma 1, Italy} 
\author{Guido Gigante*}%
\affiliation{Istituto Superiore di Sanit\`{a}, Rome, Italy}
\affiliation{Mperience s.r.l., Rome, Italy}
\author{Paolo Del Giudice}
\affiliation{Istituto Superiore di Sanit\`{a}, Rome, Italy}
\affiliation{INFN, Sezione di Roma 1, Italy} 


\begin{abstract}
The combination of new recording techniques in neuroscience and powerful inference methods recently held the promise to recover useful effective models, at the single neuron or network level, directly from observed data. 
The value of a model of course should critically depend on its ability to reproduce the dynamical behavior of the modeled system; however,  few attempts have been made to inquire into the dynamics of inferred models in neuroscience, and none, to our knowledge, at the network level.
Here we introduce a principled modification of a widely used generalized linear model (GLM), and learn its structural and dynamic parameters from \textit{ex-vivo} spiking data. We show that the new model is able to capture the most prominent features of the highly non-stationary and non-linear dynamics displayed by the biological network, where the reference GLM largely fails. 
Two ingredients turn out to be key for success. 
The first one is a bounded transfer function that makes the single neuron able to respond to its input in a saturating fashion; beyond its biological plausibility such property, by limiting the capacity of the neuron to transfer information, makes the coding more robust in the face of the highly variable network activity, and noise. 
The second ingredient is a super-Poisson spikes generative probabilistic mechanism; this feature, that accounts for the fact that observations largely undersample the network, allows the model neuron to more flexibly incorporate the observed activity fluctuations. 
Taken together, the two ingredients, without increasing complexity, allow the model to capture the key dynamic elements. 
When left free to generate its spontaneous activity, the inferred model proved able to reproduce not only the non-stationary population dynamics of the network, but also part of the fine-grained structure of the dynamics at the single neuron level.
\end{abstract}



\maketitle



Dynamic models, in neuroscience and in general, embody in a mathematical form the causal relationships between variables deemed essential to describe the system of interest; of course, the value of the model is measured both by its ability to match observations, and by its predictive power. Typically, along this route, parameters appearing in the model are assigned through a mix of insight from experiments and trial-and-error and, in turn, the study of the model dynamics helps understanding the relevance of each parameter in determining the dynamic regimes accessible to the system.  Another approach, initially quite detached from dynamic modeling, is rooted in the domain of statistical inference, and a seminal example in neuroscience was offered by the application of maximum-entropy inference of Ising-like models to multi-electrode recordings of neural activity; in this case, as well as in the later extension to kinetic Ising-like models, the Montecarlo/Glauber dynamics of the model is only meant as a means to sample the probability distribution of interest and it is not claimed to offer a detailed model of the actual dynamics at work in the system.
Recently, Generalized Linear Models (GLM) (which incorporate kinetic Ising models as a special case) have been recognized as flexible and powerful inference models \cite{truccolo2005point,truccolo2010collective}.
In time, efforts have been made to make contact between the two approaches, e.g., in the case of neuroscience, by endowing the coupling structure of the inference model with features motivated by biological plausibility \cite{capone2015inferring, pillow2008spatio, stevenson2009bayesian}; it has also been recognized that a GLM is close to a stochastic version of the spike-response model. 
Recently, the repertoire of the driven dynamics of GLM models of single neurons has been explored \cite{weber2017capturing}; however, to our knowledge, a largely open issue is to endow inference models with predictive power in terms of the system dynamics. 
Our approach to this problem is to explore the free, spontaneous dynamics of the inferred model in its relation with the one of the biological system generating the data and, in doing this, to identify the role of different elements of the inference model in determining the spontaneous dynamics of the neuronal network.
Indeed, an obvious but persisting problem in the application of inference models to neuroscience has been how to assess the meaning and value of the inferred parameters; a recurring example is offered by the inferred synaptic couplings \cite{capone2015inferring}: in the literature, a cautionary remark is usually included, recognizing that the inferred couplings (whether in the form of synaptic efficacies of more complicated synaptic kernels) are to be meant as `effective', leaving of course open the problem of what exactly this means (especially in the face of the dramatic subsampling of the underlying biological network). The value of the inferred model cannot be assessed directly by a detailed correspondence between its elements and corresponding elements of the biological system; using the output of a GLM to decode input stimuli is an interesting recent approach \cite{park2014encoding}
Our attitude is that spontaneous activity is a good testing ground to assess whether the inferred model does indeed capture essential features of the biological system, and we choose a case in which the spontaneous activity of the biological network is highly non-stationary and irregular: a cultured neuronal network generating spontaneously a wide spectrum of activity fluctuations, from population bursts, to neuronal avalanches and noisy oscillations \cite{gigante2015network,Baltz2011,eytan2006dynamics, giugliano2004single,gritsun2011experimental,Park2006a, wagenaar2006extremely, beggs2003neuronal,petermann2009spontaneous,plenz2007organizing,plenz2014criticality,lombardi2016temporal,yaghoubi2018neuronal}.
Such richness is extremely hard to reproduce with a GLM, and to our knowledge no models proposed so far were able to cope with it. For this reason we chose population bursting activity as a challenging benchmark to test our model. In our GLM approach we introduced novel ingredients inspired by biological observation, such as activity dependent negative feedback over different time-scales for the single neuron \cite{gigante2015network} (spike frequency adaptation, SFA), a bounded transfer function, and a generative probabilistic mechanism for spike generation with super-Poisson fluctuations. We show that these ingredients are crucial to endow the system with non-stationary spontaneous activity and also to account for detailed dynamic features such as temporal correlations and spatio-temporal evolution of network bursts.

\section*{Results}



\subsection*{Bounded firing rate, super-Poisson spike generation, and their impact on the spontaneous activity of the inferred model}


\begin{figure*}
\includegraphics[width=120mm]{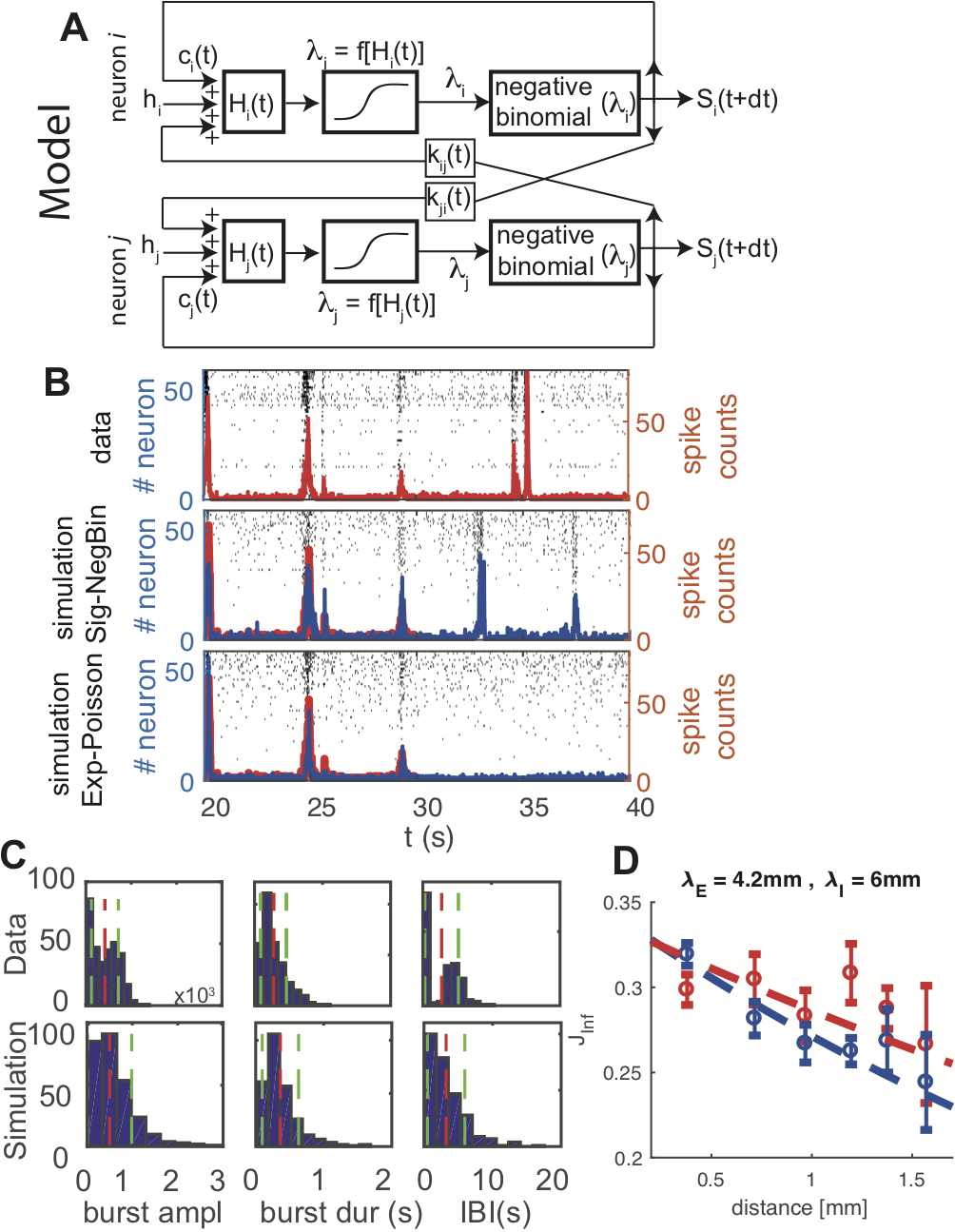} %
\caption{Spontaneous activity of the inferred model reproduces non-stationary network dynamics. 
\textbf{A} Schematic representation  of the model. The current $H_{i}(t)$ felt by neuron $i{}$ sums spikes emitted by other neurons $j$ in the recent past $(t - \Delta t)$ through synaptic kernels $k_{ij}(\Delta t)$, a constant external field $h_i{}$, and spike-frequency adaptations signals that integrate, over different time-scales, the spiking activity $S_{i}$ of neuron $i{}$ itself. This current, fed into a non-linear transfer-function $f[H_{i}]$, determines the expected number $\lambda_{i}(t)$ of spikes $S_{i}(t + \mathrm{d}t)$ that the neuron will probabilistically generate at the next time bin.
\textbf{B} Rastergrams of \textit{ex-vivo} cultured cortical neurons show a strongly non-stationary dynamics (top row; red line: whole network activity). The proposed Sig-NegBin model (central row, blue line) closely follows the data used to drive its dynamics (red line, $t$ between 20 and 30 s); when left free to run autonomously at $t = 30$ s, the model displays large irregular bursts of activity, qualitatively mimicking the behavior observed in the recordings. The standard Exp-Poisson model (bottom row, blue line), on the other hand, whilst still able to follow the data in driven mode, fails to produce any large activity fluctuation, settling instead into a noisy, stationary state.
\textbf{C} Comparison between burst amplitudes (total spike counts, left), bursts durations (center) and IBI durations (right) histograms both for the data (top row) and the autonomous activity of the Sig-NegBin model (bottom row). Mean and standard deviation of each distribution are reported with red and green lines respectively. Results for the standard Exp-Poisson model are not shown because of the extreme sparsity of burst events in that case.
\textbf{D} Average area under the inferred synaptic kernels (absolute value) as a function of the distance separating the electrodes; blue and red circles are for excitatory and inhibitory kernels respectively; dashed lines: exponential fits; the strength (area under the kernel) of inferred excitatory synapses decays faster than that of the inhibitory ones.
}
\label{fig1}
\end{figure*}

The model we introduce is an extension of GLMs as widely employed for inferring the structure of neuronal networks. With GLMs it shares the three fundamental assumptions that spike generation is a probabilistic process driven by the instantaneous current afferent to the neuron; that, secondly, such current is a linear function of the activity of the neurons in the network and, finally, that there is a non-linear transfer function linking the current to the probabilistic spike generation. The parameters of the model, in this framework, are usually obtained through the numerical maximization of the likelihood that the model itself generated the observed data. The main points of departure between more standard GLMs and the present model lay in the choice of the specific non-linear transfer function and probabilistic generative model; another notable difference is in the introduction, at the single neuron level, of an activity-dependent self-inhibition current, mimicking spike-frequency adaptation (SFA) effects widely observed in real neurons.

Fig.~\ref{fig1}, panel A, gives a schematic view of the model. The current $H_{i}(t)$ instantaneously felt by neuron $i{}$ sums spikes emitted by other neurons $j$ in the recent past $(t - \Delta t)$ through synaptic kernels $k_{ij}(\Delta t)$, a constant external field $h_i{}$, and a set of 5 SFA signals (collectively indicated as $c_{i}(t)$), that integrate, over different time-scales, the spiking activity $S_{i}$ of neuron $i{}$ itself. This current provides input to a non-linear transfer-function $f[H_{i}]$, which in turn determines the expected firing rate $\lambda_{i}(t)$ for the spike generation model at the next time bin $t + \mathrm{dt}$. The transfer function and generative model most commonly employed in the literature are, respectively, the exponential function and the Poisson distribution; in the following we will label this choice `Exp-Poisson'; our model, on the other hand, makes use of a sigmoid function (see Materials and Methods) and of a Negative-Binomial distribution with shape parameter $r_{\mathrm{NB}}$ set to $0.2$, and will be referred to in the following as `Sig-NegBin' model.

The synaptic kernels $k_{ij}(\Delta t)$, following \cite{pillow2008spatio}, are parametrized as a weighted sum of 4 `raised cosine' functions peaking at times spanning a range between 10 ms (equal to the chosen time bin $\mathrm{dt}$) and 30 ms, with support extending in the past for $15 \, \mathrm{dt} = 150 \, \mathrm{ms}$ ($k_{ij}(\Delta t) \equiv 0$ for $\Delta t > 150 \, \mathrm{ms}$). The sigmoid transfer function has two free parameters: the maximum firing rate attainable by the neuron and an asymmetry parameter that determines how fast this maximum is approached for high currents compared to the other asymptote at 0 for strongly negative currents. Finally each of the 5 SFA signals $c_{i}$, that are set equal for each neuron, has associated a characteristic time-scale $\tau_{c}$ for the integration of the spike train $S_{i}(t)$ and a weight $g_{c}$ with which $c_{i}(t)$ enters the sum of the current $H_{i}(t)$. These latter parameters, the weights defining each of the $k_{ij}(\Delta t)$ functions, and the free parameters of the sigmoid function are all determined through the likelihood maximization procedure on data  (see Materials and Methods for further details). In the case of the Exp-Poisson model, the only optimized parameters are the $k_{ij}(\Delta t)$ weights, both for $i \neq j$ and $i = j$ (`post-spike filter'); the $k_{ii}$ functions are parametrized as a weighted sum of 10 raised cosines peaking at times spanning from 10 ms to 50 ms and again extending to 150 ms in the past; the Sig-NegBin model does not include post-spike filters; thus the total number of free parameters for the Sig-NegBin is actually lower than that for the Exp-Poisson model.

Once the parameters are fitted by maximizing the likelihood, both models can be simulated to generate new spike trains according to two basic simulation modalities. The first one (`driven' mode) feeds the model with the actual spike trains observed in the data and at each time $t$ uses the model's output to make a 1-step prediction of the network activity at step $t + dt$; thus the generated activity never re-enters the network dynamics; the probabilistic distance between the 1-step prediction and the actual recorded network activity is ultimately what is minimized during the optimization procedure. In the second modality (`free' mode) the models are left free to evolve and the generated spike trains are fed back into the network, driving its dynamics at subsequent times.

We trained both the Sig-NegBin and the Exp-Poisson model on data from multi-electrode recordings of \textit{ex-vivo} cultured cortical neurons \cite{eytan2006dynamics}, that show a clear non-stationary dynamics (Fig.~\ref{fig1}B, top row) in the rastergrams and in the whole network activity (red line), with large and irregular bursts of synchronous activity interleaved by periods of low, noisy activity (`inter-burst intervals' or IBIs). At the end of the training, both models show, in driven mode, an activity that follows quite closely the experimental spike trains (Fig.~\ref{fig1}B, middle and bottom rows; red line: data; blue line and rastergrams: simulation). However, when the simulation is switched to free mode at time t = 30 s, the behavior of the two models clearly diverges: whilst the Sig-NegBin is able to autonomously sustain a highly non-stationary activity, the Exp-Poisson only exhibits stationary fluctuations. Note that, in free mode, the activity of the Sig-NegBin model does not closely follow the activity of the real network anymore (top row, last 10 s); this is not surprising: the choice of a probabilistic model as a GLM stems from the assumption that the network dynamics is, to a large extent, influenced by self-induced random fluctuations and, thus, non-deterministic; it is therefore expected that, at least after a transient phase, the model and the data will drift away. Yet the activity spontaneously generated by the model clearly resembles the one seen in data in an intuitive statistical way, showing irregular bursts interspersed with interval of relative quiescence.

Such resemblance is made more quantitative in Fig.~\ref{fig1}C. Bursts of the network global activity are detected through the algorithm developed in \cite{gigante2015network}, both for the data (top row) and the autonomous activity of the Sig-NegBin model (bottom row). The comparison between burst amplitudes (total spike counts, left), bursts durations (center) and IBI durations (right) histograms shows a semi-quantitative agreement between the dynamics of the biological network and the inferred model (mean and standard deviation of each distribution are reported with red and green lines respectively). The main discrepancy between model and data is visible in the IBI distribution; ``doublets'' of bursts in the data, clearly separated by the rest of the IBI distribution, are absent in the model. Based on previous modeling work on similar data \cite{masquelier2013network}, the discrepancy is likely due to elements, we did not include in our minimal model, such as short-term synaptic facilitation and depression.

We remark that the analogous distributions for the Exp-Poisson model are not reported because of the extreme sparsity of burst events in that case. Moreover we observe that, despite the major difference in the exhibited dynamic behavior, the inferred synaptic kernels for our model and the Exp-Poisson model are correlated ($c_{\mathrm{Pearson}} = 0.49$ with a $P<0.05$; correlation between the total areas under each synaptic kernel, as a measure of `synaptic strength'). The inferred synaptic kernels also show a correlation with the distance separating the electrodes where the activity of the pre- and post-synaptic neurons where recorded (Fig.~\ref{fig1}D; blue and red circles are for excitatory and inhibitory kernels respectively; dashed lines: exponential fits). It is worth noting that the connections' decay with distance is faster for excitatory than for  inhibitory inferred synapses, consistently with what is found in activity correlations in the neocortex by using 2D multielectrode arrays \cite{peyrache2012spatiotemporal}.

\begin{figure*}
\includegraphics[width=130mm]{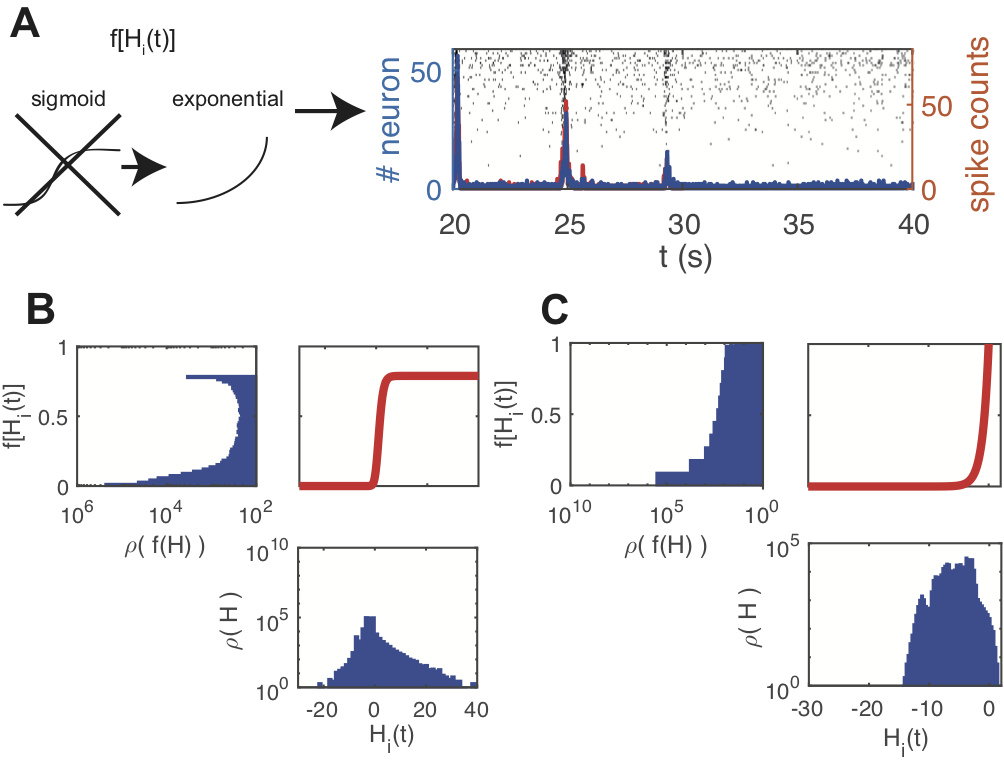} %
\caption{%
Bounded spike rate is necessary to generate non-stationary spontaneous activity in the inferred model.
\textbf{A} Population activity (blue line) and rastergram generated by the model where the bounded transfer function is replaced by an exponential and inference procedure is carried out from scratch. Red line represent the data driving the dynamics for the first 10 s ($20 \leq t \leq 30$ s); afterwards, when left free to generate its spontaneous activity, the model is unable to sustain any non-stationary bursting activity and settles in a quiescent state.
\textbf{B-C} transfer function (respectively sigmoid and exponential) $f[\cdot]$ (top-right panel) distribution of input currents $H_{i}(t)$ to the single neuron across different times (bottom panel) and distribution of firing rates $\lambda_{i}(t) = f[H_{i}(t)]$ ( top-left panel) 
}
\label{fig2}
\end{figure*}

The choice of a sigmoidal, saturating transfer function, besides seeming just natural in a GLM meant to model neural activities, turns out to be key in achieving the above results. We repeated the inference procedure with a hybrid Sig-NegBin model, in which the sigmoidal transfer function is replaced with an exponential (Fig.~\ref{fig2}A). As above, when during the first 10 s the model is simulated (blue line and rastergrams) in driven mode, it shows a good agreement with the driving data (red line); when the driving is turned off, instead, the model settles in a quiescent state, absent any large bursts of activity. In free mode, the saturation of the sigmoidal function $f[\cdot]$ (Fig.~\ref{fig2}B, top-right panel) allows a wide distribution of input currents $H_{i}(t)$ to the single neuron across different times (Fig.~\ref{fig2}B, bottom panel) to give rise to a distribution of firing rates $\lambda_{i}(t) = f[H_{i}(t)]$ (Fig.~\ref{fig2}B, top-left panel) that is strongly bimodal (notice the logarithmic scale on the y-axis).
On the other hand, the fast growth of the exponential function (Fig.~\ref{fig2}C, top-right panel) pushes the inference procedure towards somewhat narrower, more conservative distribution of currents (Fig.~\ref{fig2}C, bottom panel), resulting in a unimodal distribution of firing rates. The comparison of the two models in driven and free modes suggests that the exponential transfer function makes the system very selectively susceptible to the fluctuations experienced during learning, while the sigmoid transfer function allows for susceptibility to a wider range of fluctuations, which we believe is the root of the ability to spontaneously generate network bursts.



\begin{figure*}
\includegraphics[width=120mm]{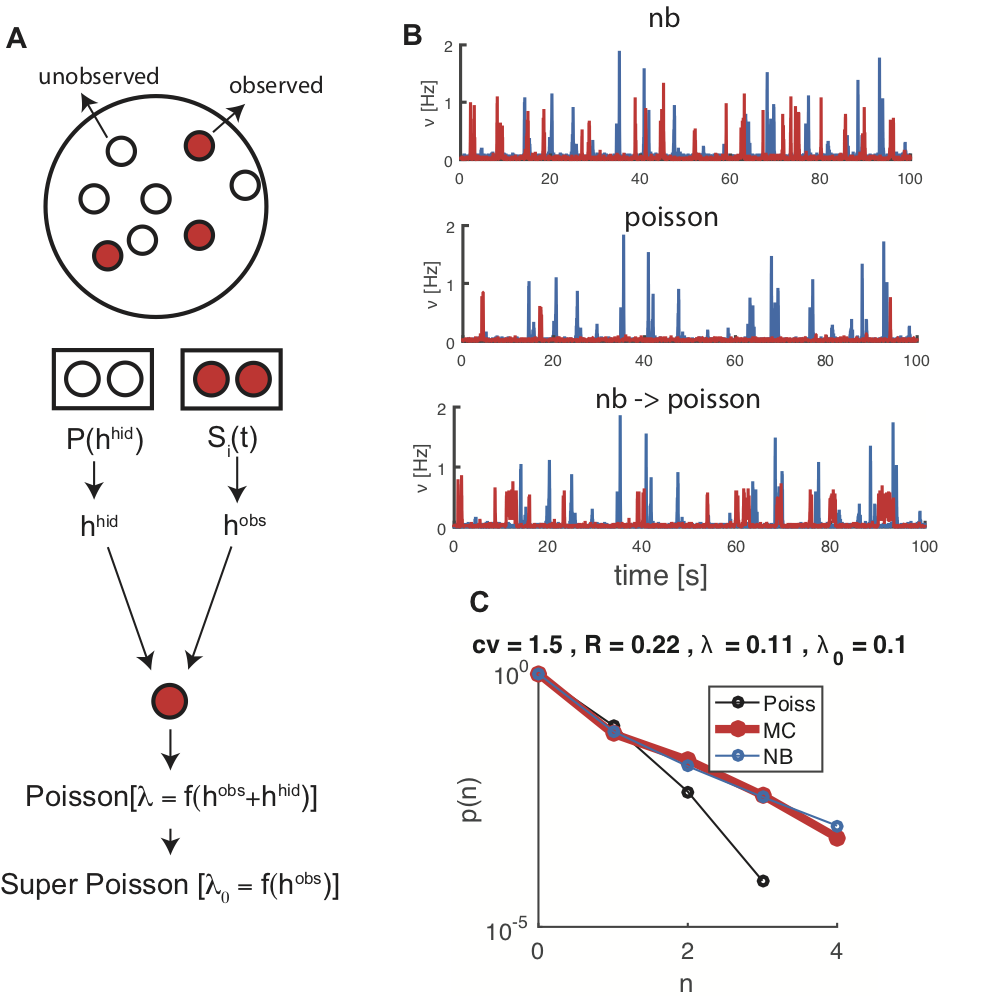} %
\caption{%
Rationale and implications of choosing a negative binomial as generative model of spike counts.
\textbf{A} Sketch showing the effects of the unobserved neurons on the fluctuations in the observed part of the network.
\textbf{B} Spontaneous bursting activity generated by different models (blue line: data; red line: model in free mode). Upper panel: Sig-NegBin model; middle panel: Sig-Poisson; bottom panel: Sig-NegBin model after replacing the Negative Binomial with a Poisson spike generator.
\textbf{C} Comparison between the spike counts generated by a negative binomial distribution ($r = 0.22$) with a mean value equal to the average spike count of the neuron (blue line), the Monte Carlo sampled $p(S_{i}| H_{i})$ with best matching values for the first two moments of spike counts (red line) and Poisson distribution for the same average spike count (black line).
}
\label{fig3}
\end{figure*}

The recorded neurons from the cultured network constitute of course a dramatic subsampling of the network (Fig.~\ref{fig3}A, leftmost sketch; though, in fact, the activity of several neurons will be collected by each electrode). The effects of the unobserved neurons on the behavior of the observed part of the network could be in principle very complex \cite{levina2017subsampling}; yet we argue that the adoption of a Negative Binomial distribution for the spike counts generation, instead of a Poisson distribution as commonly assumed, captures a relevant part of those effects and is in fact critical to account for the non-stationarity of the spontaneous activity of the network.

One way to approach the problem is to assume that at each time $t$ the generated spike count of a generic neuron $S_{i}(t)$ is determined by both the field $H_{i}(t) = h_{i} + \sum_{\tau}  \sum_{j} \alpha_{ij}(\tau) \, S_j(t - \tau)$ (the sum of external fields and the one due to the activity of the other observed neurons), and the field due to inputs from unobserved, 'hidden' neurons, $h_{i}^{\mathrm{hid}}$ (Fig.~\ref{fig3}A), such that the observed spike count probability distribution would be obtained by integrating over the (unknown) distribution of $h_{i}^{\mathrm{hid}}$:
\begin{equation}
p(S_{i}| H_{i}) = \int d h_{i}^{\mathrm{hid}} p(S_{i}| H_{i}, h_{i}^{\mathrm{hid}}) \,\, p(h_{i}^{\mathrm{hid}})
\label{spikeCountDistrib}
\end{equation}
where we still assume that $p(S_{i}(t)| H_{i}(t), h^{\mathrm{hid}})$ is a Poisson distribution, with mean given by $\lambda_{i}(t) = f\big[H_{i}(t) + h_{i}^{\mathrm{hid}}\big]$.

It is intuitive that $p(S_{i}(t)| H_{i}(t))$ will now generate super-Poissonian fluctuations; and it turns out that larger fluctuations are critical for the ability of the Sig-NegBin model to spontaneously generate bursting activity, as shown above (Fig.~\ref{fig3}B upper panel; blue line: data; red line: Sig-NegBin model in free mode). In fact, an inferred Sig-Poisson model shows very sparse bursting (Fig.~\ref{fig3}B, middle panel). But it is not just that a broader spike generation distribution favors larger fluctuations in the network activity. We simulated the Sig-NegBin model replacing the Negative Binomial with a Poisson spike generator (Fig.~\ref{fig3}B, bottom panel); such `Poisson replay' of the inferred Sig-NegBin model generates frequent bursts, each lasting for long time on average. Thus the inferred synaptic couplings of the Sig-NegBin model, unlike the ones for the Sig-Poisson, appear to be compatible with a bistable network dynamics, between UP and DOWN states, as also suggested by the results for the sigmoid transfer function (Fig.~\ref{fig2}B). Therefore, the main role of a super-Poisson spike generation distribution seems to be to destabilize the UP state, making the inferred model once again more robust to the variability of the free network's activity and thus able to self-sustain a highly non-stationary activity.

Taking together the results in Fig.~\ref{fig2} and Fig.~\ref{fig3}, the suggested picture is that on the one hand the sigmoid gain function allows the inference procedure to explore and use ranges of couplings that are effectively forbidden for an exponential gain function, and are essential to spontaneously generate large sudden increase of activity (burst onset); on the other hand, the NegBin generator plays the dynamic role of efficiently destabilizing, with large fluctuations of the spike counts, the high-activity states which are spontaneously generated by the large recurrent excitation. The two ingredients are robustly coupled to spontaneously produce bursting activity.

Albeit representing just one of the forms that $p(S_{i}(t)| H_{i}(t))$ can take, the NegBin distribution has been suggested to adequately model fluctuations in observed neural activity \cite{onken2009analyzing, goris2014partitioning}, which makes it a natural candidate. Yet, a NegBin would result from Eq.~\ref{spikeCountDistrib} assuming an exponential transfer function and a log-Gamma distribution for $h_{i}^{\mathrm{hid}}$: the first condition is patently in contradiction with our choice of a saturating transfer function and the second one is not expected to hold in general. We nevertheless tested the adequacy of the NegBin assumption for the Sig-NegBin model as follows. We assumed that $p(h_{i}^{\mathrm{hid}})$ in Eq.~\ref{spikeCountDistrib} is a scaled and shifted version of the distribution of the observed fields $H$; in turn we estimated such distribution from the inferred synaptic couplings and the observed experimental spike counts, for the neuron with maximal average activity, for which we expected the differences in the spike count distribution would both show up more clearly, and matter more for the dynamics. From this, we performed a Monte Carlo estimate of $p(S_{i}| H_{i})$ for different values of the mean and the standard deviation defining $p(h_{i}^{\mathrm{hid}})$. Fig.~\ref{fig3}C shows a NegBin that is very close to the one used in the Sig-NegBin model ($r = 0.22$) for a mean value equal to the average spike counts of the neuron (blue line); the sampled $p(S_{i}| H_{i})$ with best matching values for the first two moments is reported in red; the two distributions are very similar, especially if compared to the Poisson distribution for the same average spike count (black line). Thus spike counts compatible with a NegBin distribution are naturally accounted for by a plausible assumption on the effect of sub-sampling underlying Eq.~\ref{spikeCountDistrib}, with marked deviations from a Poisson distribution. We repeated the above procedure sampling $h_{i}^{\mathrm{hid}}$ from a Gaussian distribution, finding similar results; this suggests that, provided suitable values for the mean and variance are chosen, the precise shape of the $p(h_{i}^{\mathrm{hid}})$ has a minor effect.

\subsection*{The inferred model captures detailed temporal and spatial aspects of the neural dynamics}

\begin{figure*}
\centering
\includegraphics[width=150mm]{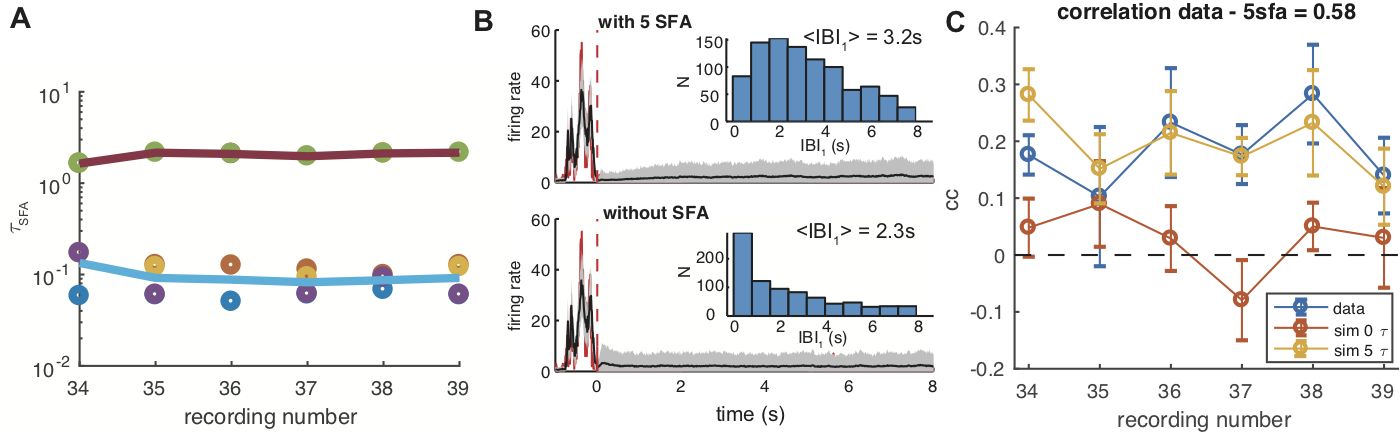} %
\caption{Effect of inferred time-scales and SFA currents.
\textbf{A} Inferred $\tau_{\mathrm{SFA}}$ for different recordings for the same preparation; the five inferred time-scales cluster around two values for all recordings, one about $100 ms$, the other around $2s$. 
\textbf{B} Comparison between the model dynamics with and without SFA. In black, and gray-shading, are represented the average and variance over 500 simulations. Red line is the average activity from data during a large chosen burst. Insets: histograms of the time to the subsequent burst.
\textbf{C} Correlation between the burst amplitude and the time since the preceding burst.}
\label{fig4}
\end{figure*}

Although the spike-frequency mechanism with which the Sig-NegBin model is endowed allows for 5 independent time-scales, it is interesting to note that the inferred values consistently aggregate around just two values, at about $100 \, \mathrm{ms}$ and $2 \, \mathrm{s}$ respectively, for different temporal segments of the same network's activity (Fig.~\ref{fig4}A). This result is consistent with previous work on the same data aiming to infer the main time-scales of the bursting dynamics via a completely different approach \cite{gigante2015network}.

To assess how relevant the adaptation mechanism is for the model dynamics, we performed inference with and without SFA. Then we simulated the model in driven mode until just after the end of a burst of large amplitude, where adaptation effects are expected to be more pronounced, leaving the model free to spontaneously evolve afterwards. In Fig.\ref{fig4}B the SFA and non-SFA dynamics are reported (top and bottom, respectively); black line is the model activity averaged over 500 simulations of the dynamics, while gray shading marks the plus/minus one standard deviation range; red line is the population activity from data. For each realization we collected the next IBI, that is the time interval to the next burst spontaneously generated by the model. The IBI histogram (inset) attains a maximum very close to zero when SFA is not present; such maximum shifts toward higher values with SFA; the two distributions also have different averages ($2.3$ s \textit{vs} $3.2$ s) and coefficients of variation ($0.98$ \textit{vs} $0.65$). Thus SFA significantly increases the average interval to the next burst, reducing at the same time its variability.

We also found in data, consistently with the results reported in \cite{lombardi2016temporal}, a positive correlation between burst amplitude and the length of the previous IBI, an effect that could be possibly attributed to some adaptation mechanism, such as SFA. We estimated this correlation from simulations of the models inferred, with and without SFA, on different recordings; Fig.~\ref{fig4}C compares the result of the two models with that found in their experimental counterparts. It is seen that the model without SFA is unable to recover correlations that are significantly different from zero, while the model with SFA produces correlations that are consistent with that measured in the data.



\begin{figure*}
\includegraphics[width=120mm]{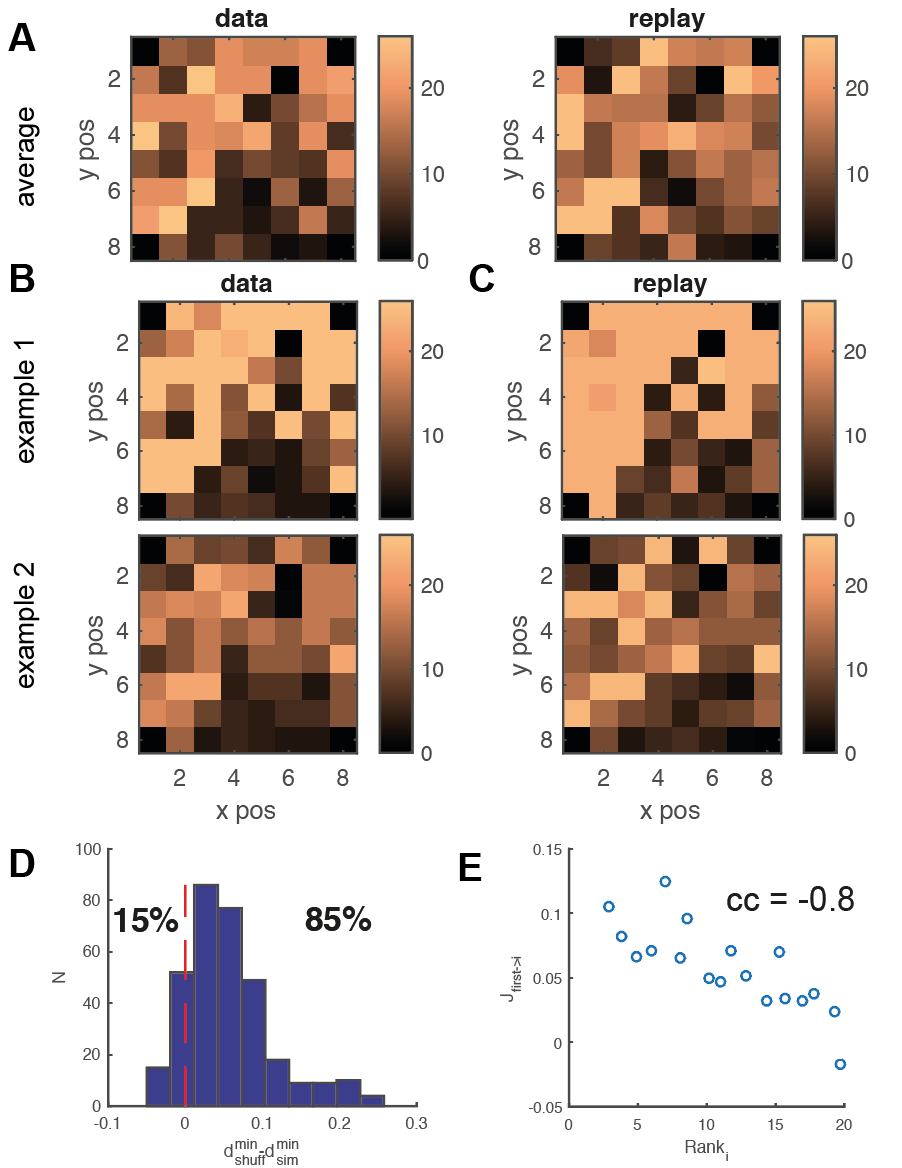} %
\caption{%
Spatial microscopic network dynamics, data \textit{vs} model. 
\textbf{A} Spatial distribution of the average time-rank of neuron activations during bursts for data and model. Averages are computed over all bursts in the data and a simulation of comparable length.
\textbf{B} Spatial distribution of time-rank for 2 single bursts in data. 
\textbf{C} Spatial distribution of time-rank for the 2 bursts from simulation, most similar to those showed in B. 
\textbf{D} Distribution of the difference of the similarity between bursts from data, and both bursts from simulation and bursts from surrogate simulated data.
\textbf{E} Scatterplot of post-synaptic efficacies of the neuron with smallest time-rank \textit{vs} the rank of its post-synaptic targets.}
\label{fig5}
\end{figure*}

We also addressed the ability of the model to account for the temporal structure and spatial organization of the network activity. We first considered the average order of activation of different spatial locations within a burst (see Methods for details): time-rank is the index of the time bin when a neuron spikes first since the burst onset. By evaluating the average rank for both each neuron in the model and each electrode in the data, we found them strongly correlated ($c = 0.9$). In Fig.~\ref{fig5}A we report the spatial distribution of average time-ranks for data (left) and simulations (right).

Looking farther into the burst temporal structure, we asked to what extent the time-ranks in the model and in the data are comparable on a single burst basis. For each burst in the data, we selected and compare the simulated burst with the closest time-rank pattern. Fig~\ref{fig5}B and C show two examples of such comparison. To provide a global comparison, capturing the correlation in time-ranks between data and simulations, we proceeded as follows: independently for each neuron, we shuffled the vector of its time-ranks across all the simulated bursts, thereby destroying spatial correlations in the time-ranks, while preserving the average ranks (Fig.~\ref{fig5}A, right).

For each burst in the data we took the closest simulated burst and the closest shuffled simulated one, computed the corresponding distances ($d_{shuff}^{min}$ and $d_{sim}^{min}$) and the distribution of their differences (Fig.~\ref{fig5}D). It is seen that, for the large majority of bursts, the simulated burst is closer to the data burst than the surrogate. We therefore can conclude that inferred model captures most of the spatial development of neural activity at the single burst level.

To inquire into the relationship between the inferred synaptic structure and the ensuing network dynamics, we also asked whether for a neuron with low average time-rank (early spiking), the efficacy of its outgoing synapses correlates with the time-ranks of its post-synaptic neurons. Fig.~\ref{fig5}E shows indeed a high negative correlation between the efficacy of outgoing synapses and the time-rank of post-synaptic neurons ($c = -0.8$).

\section*{Discussion}
Several criteria can be identified to evaluate a model, yet the ability to reproduce the behavior of the system under analysis is certainly among the most relevant. We have seen that the addition of few computational ingredients can enable a probabilistic, generative network to autonomously sustain a rich, highly non-stationary dynamics, as observed in the experimental data employed to infer the model's parameters.

In recent years, the field of statistical inference applied to neuronal dynamics has mainly focused on devising models and procedures that could reliably recover the real or effective synaptic structure of a network of neurons, under different dynamical and stimulation conditions \cite{roudi2015multi,tyrcha2013effect,nghiem2018maximum}, even though, more often than not, an experimental ground truth was not readily available.

Although, quite surprisingly, the study of the dynamics of the inferred models has been largely neglected, we are not the first to address the issue. In \cite{pillow2008spatio} the authors demonstrated, on an in-vitro population of ON and OFF parasol ganglion cells, the ability of a GLM to accurately reproduce the dynamics of the network; \cite{park2014encoding} studied the response properties of lateral intraparietal area neurons at the single trial, single cell level; the capability of GLM to capture a broad range of single neuron response behaviors was analyzed in \cite{weber2017capturing}. In all these works, however, the focus was on the response of neurons to stimuli of different spatio-temporal complexity; even where network interactions were accounted for, they proved to be, for the overall dynamics displayed by the ensemble, an important, yet not decisive correction. To our knowledge, no published study to date has focused on the autonomous dynamics of GLM networks inferred from neuronal data.

Important progress has also been made on the ability of GLMs to predict single neuron spiking in an ensemble of neurons, with accuracy higher than that provided by methods based on the instantaneous state of the ensemble itself \cite{truccolo2010collective} and the PSTH \cite{pillow2008spatio}. In the lingo of the present paper, though, these attempts were based on simulations of the inferred network in driven mode, with at most the activity of a single neuron free to reenter just that single neuron's dynamics. To our knowledge, our work for the first time has attempted a direct, microscopic comparison between the activity autonomously generated by an inferred GLM network and the one observed in the biological data. The free mode of simulation opens the way, in principle, to multi-time-step, whole network activity predictions. Being the model probabilistic and the data noisy, it is of course expected that the ability of the model, in free simulation mode after having been driven by the data for some time, to follow the experimentally observed spike trains should quickly deteriorate. In fact, this is what we found with our model: a burst becomes practically unforeseeable going some 100-200 ms in the past. This result is in contrast with the impressive findings reported in \cite{tajima2017locally}, in which the authors, on recordings similar to the ones used in this work, through a new model-free method based on state-space reconstruction, were able to predict the occurrence of a burst even 1 s in advance. Whilst it is not clear why this is the case, we applied the state-reconstruction method on our data with negative results, in agreement with what we found with our model.

We presented several semi-quantitative comparisons between the spike trains generated by the model network, without any time-varying external input, and the data from which the model's parameters have been inferred. Such tests, albeit successful, represent of course just an example of the many possible ways to compare two bursting dynamics. Yet we stress how a similar model, that has been widely and successfully applied to the analysis of neuronal data and has a slightly greater number of parameters too, failed in large part to reproduce the most prominent features of the dynamics displayed by the data.

Two interrelated questions pertain to how representative the chosen experimental system is of neuronal network dynamics in general and the capability of the presented framework to generalize to other kinds of non-stationary neural activities. As for the first question, our primary concern in selecting an \textit{ex-vivo} bursting culture has been to study a system that, absent any external stimuli, would show a non-trivial collective behavior. As discussed above, the capability of GLMs to mimic the response properties of single neurons or a network to stimuli has already been well investigated. On the other hand, numerous studies in literature witness to the complexity of spatial and temporal behavior of neuronal cultures \cite{gigante2015network,Baltz2011,eytan2006dynamics, giugliano2004single,gritsun2011experimental,Park2006a, wagenaar2006extremely, beggs2003neuronal,petermann2009spontaneous,plenz2007organizing,plenz2014criticality,lombardi2016temporal,yaghoubi2018neuronal}. As for the second question, we remark that our primary goal was to devise a `proof of concept' extension to show that a GLM can autonomously generate a complex non-stationary activity, and not to establish general conditions for such capability. We therefore chose a bursting regime as a testbed for the model not because it is exemplary, but because it represented a clear challenge: a collective, non-linear phenomenon that demands the model to adapt to a broad spectrum of time- and space-scales. There were of course sensible alternatives to choose from, the most relevant being, \textit{e.g.}, the UP and DOWN dynamics in brain slices; although we have not tackled the issue directly, our results (see Fig.~\ref{fig3}B, bottom plot) hint at a inherent capability of the model to sustain a noisy bistable dynamics, as more in general suggested in \cite{rostami2017bistability}. Nevertheless, the extent to which generative models are able to adapt to different regimes, and how dynamic network effects and stimulation characteristics interact in a generative model like the one studied here are open issues, that will probably attract much interest in the near future.

Given our goal to establish the capability of a GLM to reproduce a complex, non-linear dynamics, we found two ingredients - a non-symmetric sigmoidal transfer function and a Negative Binomial spike-generation model - that proved to be minimal: absent one of them, the model's performances were drastically impaired. Even if we do not claim these specific ingredients to be necessary in general, we are persuaded they represent instances of mechanisms that are both biologically plausible and computationally desirable for the model.
Although a sigmoidal, saturating transfer function would appear to be a natural choice for a model meant to reproduce neural activities, surprisingly this option has never been explored, to our knowledge, to perform network inference on neuronal recordings. While the use of an exponential transfer function in GLMs is grounded in general statistical requirements \cite{lindsey2000applying}, in our case an asymmetric sigmoid appears to be the single most important factor explaining the success of the proposed model. As already noted, the saturation of the sigmoidal function allows naturally for a bimodal distribution of firing rates and thus makes model's behavior more robust in the face of the intrinsic fluctuations of network activity. Therefore, if the specific form we chose is just one possible instance among many plausible ones, a saturating behavior is expected to benefit the model beyond the immediate scope of the present work; a certain degree of asymmetry also appears to be beneficial according to our results, allowing for differential sensitivity to high and low input levels.


It is interesting to note how this finding contrast with the success of non-saturating transfer functions in deep learning literature, where the introduction of rectified linear units, in place of the standard logistic ones, has represented one of the major breakthroughs of recent years \cite{nair2010rectified}. Such units exhibit `intensity equivariance', that is the ability to readily generalize to data points that differ only for a scale factor; whilst such property is clearly valuable when dealing with data such natural images and sounds, when applying machine learning techniques to very noise and sparse data, such as in the case studied here, bounded transfer functions are probably beneficial exactly for the opposite reason: they filter out most of the incoming information to gain a poorer but more robust coding in the output. 


We have seen that, if the role of a sigmoid gain function seems to facilitate a bistable behavior of the network, the negative binomial's one is to efficiently destabilize, with large fluctuations of the spike counts, the high-activity state. Recent experimental evidence has emerged supporting the Negative Binomial distribution as a candidate for the spike counts variability in real neurons \cite{goris2014partitioning}; moreover, a negative binomial spike generation has already been adopted in a model combining GLM with flexible graph-theoretic priors for the connectivity \cite{linderman2016bayesian}. Although, then, our second ingredient already finds experimental and theoretical support in the literature, we provide a new hypothesis on why the super-Poisson statistics arises: as the effect of input fluctuations generated by the activity of neurons that have not been recorded. And our hypothesis directly hints to the Negative Binomial as just one possible way to model such effect, where over-dispersed spike counts are, instead, a necessary signature of it; it is this more general feature, in our opinion, that proved to be so important for the success of the proposed model. Interestingly, our findings resonate with the recently proposed role of fluctuating unobserved variables in the emergence of criticality in a wide range of systems \cite{schwab2014zipf}. 

Our hypothesis presupposes that the recorded neurons have been chosen at random from the whole network; this is probably not the case in many instances, where a local set of units is sampled instead; such non-random sub-sampling can potentially lead to strong deviations from expected behavior \cite{levina2017subsampling}; this of course could produce systematic effects in the inference. Besides, our model cannot incorporate the effects of slow changes in single neuron excitability, as observed experimentally in \cite{gal2010dynamics, gal2013self}, and analyzed in a statistical model in \cite{goris2014partitioning}.


\section*{Methods}

\subsection*{Experimental data}
As originally described in \cite{eytan2006dynamics}, cortical neurons were obtained from newborn rats within 24 hours after birth, following standard procedures. Briefly, the neurons were plated directly onto a substrate-integrated multielectrode array (MEA). The cells were bathed in MEM supplemented with heat-inactivated horse serum (5\%), glutamine (0.5 mM), glucose (20 mM), and gentamycin (10 $\mu$g/ml) and were maintained in an atmosphere of $37^{\circ}$ C, 5\% CO2/95\% air in a tissue culture incubator as well as during the recording phases. The data analyzed here was collected during the third week after plating, thus allowing functional and structural maturation of the neurons. MEAs of 60 Ti/Au/TiN electrodes, 30 $\mu{}$m in diameter, and spaced 200 $\mu$m from each other (Multi Channel Systems, Reutlingen, Germany) were used. The insulation layer (silicon nitride) was pretreated with poly-D-lysine. All experiments were conducted under a slow perfusion system with perfusion rates of $\sim$100 $\mu$l/h. A commercial 60-channel amplifier (B-MEA-1060; Multi Channel Systems) with frequency limits of 1-5000 Hz and a gain of 1024$\times$ was used. The B-MEA-1060 was connected to MCPPlus variable gain filter amplifiers (Alpha Omega, Nazareth, Israel) for additional amplification. Data was digitized using two parallel 5200a/526 analog-to-digital boards (Microstar Laboratories, Bellevue, WA). Each channel was sampled at a frequency of 24000 Hz and prepared for analysis using the AlphaMap interface (Alpha Omega). Thresholds (8$\times$ root mean square units; typically in the range of 10-20 $\mu V$) were defined separately for each of the recording channels before the beginning of the experiment. The electrophysiological data are freely available from S. Marom (\url{http://marom.net.technion.ac.il/}).





\section*{Acknowledgements} We are grateful to Shimon Marom for having shared his data, and for several illuminating discussions along the way.




%

\end{document}